\documentclass{emulateapj}

\newcommand{\hi}{\mbox{\rm \ion{H}{1}}}

\newcommand{\htwo}{\mbox{\rm H$_2$}}

\shorttitle{Tightly Correlated {\sc HI} and FUV Emission in the Outskirts of M83}
\shortauthors{Bigiel et al.}

\begin{document}

\slugcomment{Accepted for Publication in ApjL}
\title{Tightly Correlated {\sc HI} and FUV Emission in the Outskirts of M83}

\author{F.~Bigiel\altaffilmark{1,2}, A.~Leroy\altaffilmark{3,6},
M.~Seibert\altaffilmark{4}, F.~Walter\altaffilmark{2}, L.~Blitz\altaffilmark{1},
D.~Thilker\altaffilmark{5}, B.~Madore\altaffilmark{4}}

\altaffiltext{1}{Department of Astronomy, Radio Astronomy Laboratory,
University of California, Berkeley, CA 94720, USA; bigiel@astro.berkeley.edu}
\altaffiltext{2}{Max-Planck-Institut f{\"u}r Astronomie, K{\"o}nigstuhl 17,
69117 Heidelberg, Germany}
\altaffiltext{3}{National Radio Astronomy Observatory, 520 Edgemont Road, Charlottesville,
VA 22903, USA}
\altaffiltext{4}{Observatories of the Carnegie Institution of Washington,
  Pasadena, CA 91101, USA}
\altaffiltext{5}{Center for Astrophysical Sciences, Johns-Hopkins University,
3400 North Charles Street, Baltimore, MD 21218, USA}
\altaffiltext{6} {Hubble Fellow}

\begin{abstract}
We compare sensitive \hi\ data from The \hi\ Nearby Galaxy Survey
(THINGS) and deep far UV (FUV) data from GALEX in the outer disk of
M83. The FUV and \hi\ maps show a stunning spatial correlation out to
almost 4 optical radii (${\rm r_{25}}$), roughly the extent of our
maps. This underscores that \hi\ traces the gas reservoir for outer
disk star formation and it implies that massive (at least low level)
star formation proceeds almost everywhere \hi\ is observed. Whereas
the average FUV intensity decreases steadily with increasing radius
before leveling off at $\sim 1.7{\rm r_{25}}$, the decline in
\hi\ surface density is more subtle. Low \hi\ columns
($\lesssim2\,{\rm M}_{\odot}~{\rm pc}^{-2}$) contribute most of the
mass in the outer disk, which is not the case within $r_{25}$. The
time for star formation to consume the available \hi, inferred from
the ratio of \hi\ to FUV intensity, rises with increasing radius
before leveling off at $\sim100$\,Gyr, i.e., many Hubble times, near
$\sim 1.7{\rm r_{25}}$. Assuming the relatively short \htwo\ depletion
times observed in the inner parts of galaxies hold in outer disks, the
conversion of \hi\ into bound, molecular clouds seems to limit star
formation in outer galaxy disks. The long consumption times suggest
that most of the extended \hi\ observed in M83 will not be consumed by
{\em in situ} star formation. However, even these low star formation
rates are enough to expect moderate chemical enrichment in a closed
outer disk.
\end{abstract}


\section{Introduction}

In this letter we compare wide-field images of atomic hydrogen (\hi)
and far ultraviolet (FUV) emission, a tracer of recent star formation (SF),
in the far outer disk of the nearby spiral galaxy M83. Star formation in
the outer disks of galaxies has been the subject of numerous studies,
including direct optical observations of massive SF
\citep[e.g.,][]{ferguson98, lelievre00, cuillandre01, deblok03} and
studies of the star-forming interstellar medium (ISM) using CO
emission \citep{braine07,gardan07} or dust
\citep{zaritsky94,popescu03}. With its large field-of-view and
excellent sensitivity, the {\em Galaxy Evolution Explorer} (GALEX)
revolutionized this field, revealing widespread, extended SF in the
far outer reaches (i.e., far beyond the optical disks) of many
galaxies
\citep{thilker05,thilker07,thilker09,gildepaz05,gildepaz07a,boissier07}. At
the same time it has long been known that many galaxies host extended
\hi\ distributions \citep[e.g.,][] {bosma81,kamphuis92, bajaja94,
  boomsma08,walter08} -- including M83 \citep[e.g.,][]{huchtmeier81}.

Mainly due to the lack of matched wide-field and sensitive UV and 21\,cm
observations, it is still largely unclear how these two
extended components -- star formation traced by FUV emission and
atomic hydrogen -- relate to one another. Although the integrated star
formation rates (SFRs) in extended UV (XUV) disks represent only a small
fraction of the total SFR, the \hi-SFR connection at large radii
bears on a number of aspects of galaxy structure and evolution, such as
the shapes and edges of stellar disks \citep{pohlen06} or chemical
enrichment gradients across galaxies \citep{gildepaz07b}. The
consumption of outer disk \hi\ by SF affects the availability of this
gas for fueling SF in inner galaxy disks, a necessary process for
galaxies to sustain SF over cosmological times
\citep[e.g.,][]{shlosman89,blitz96,bauermeister09}. Finally, comparing
UV and \hi\ in the extreme (low-density, often low-metallicity)
environment of outer galaxy disks can illuminate the limiting
conditions for cloud and star formation (``star formation
threshold").

In this letter we combine sensitive, large field-of-view
\hi\ data from THINGS \citep[The \hi\ Nearby Galaxy
  Survey,][]{walter08} with extremely deep FUV data from GALEX to
study the relationship between \hi\ (the dominant component of the ISM
at large radii) and FUV emission in the nearby spiral galaxy M83. Previous, shallower GALEX
observations showed that M83 hosts one of the most prominent examples
of a XUV disk \citep{thilker05,thilker07}. This unique
data set allows us to trace the distribution of \hi\ and FUV emission
across a $\sim50\arcmin$ field-of-view, corresponding to almost 4
optical radii r$_{25}$ (defined as the 25th B-band mag
arcsec$^{-2}$ isophote). We use these data to measure the location
(\S\,\ref{sec:location}) and amount (\S\,\ref{sec:amount}) of FUV
emission relative to the \hi. In Section \S\,\ref{dist-hi} we
investigate the distribution of \hi\ surface densities in different radial
regimes across M83.

\begin{figure*}
\plotone{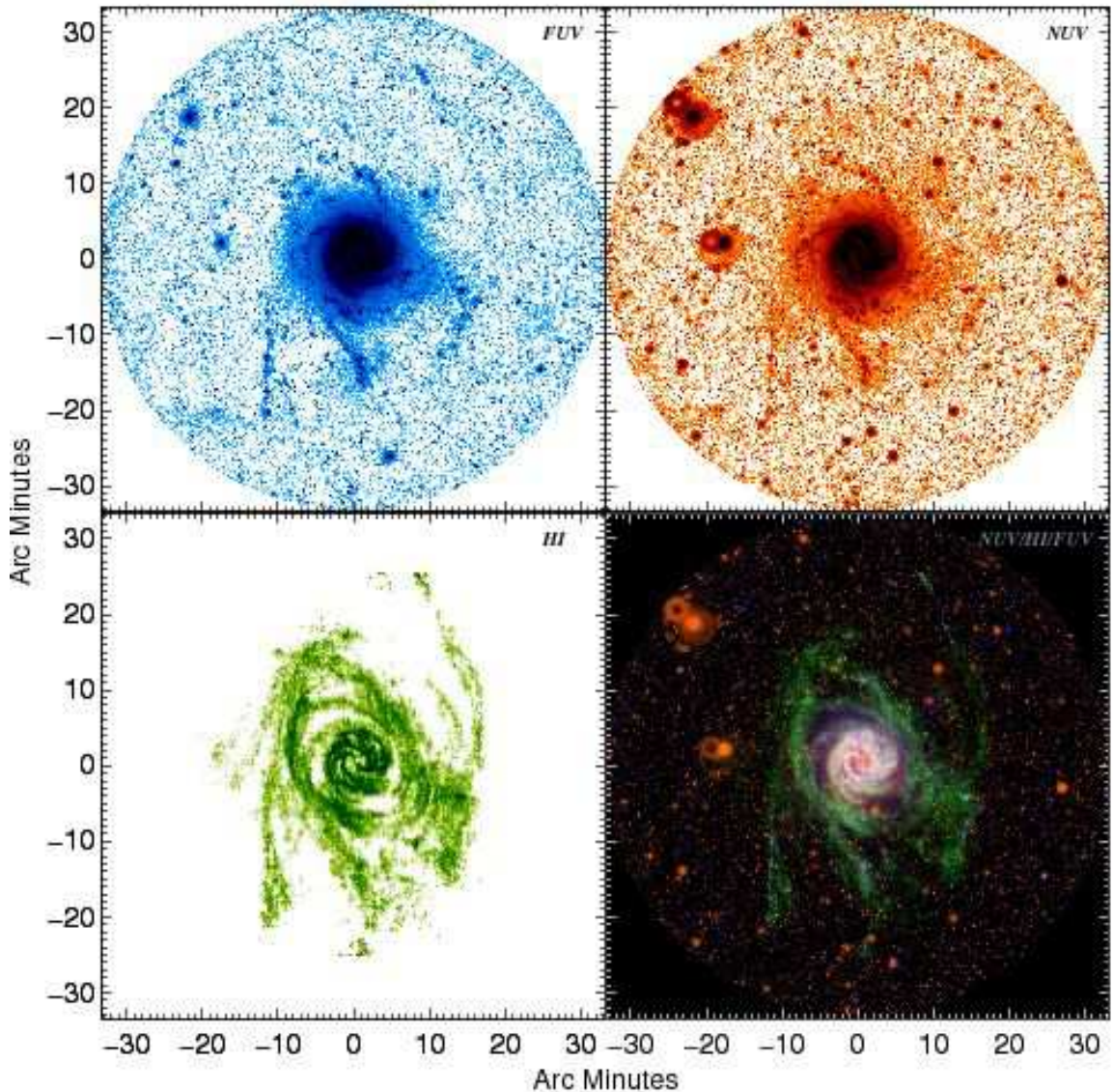}
\caption{GALEX far UV (upper left), near UV (upper right), THINGS {\sc
    Hi} (lower left) maps of M83 and a combination of smoothed {\sc
    Hi}, near, and far UV (lower right).  The axes show the offset
  from the field center at ${\rm R.A.=13^{h}37^{m}1.2^{s}}$ and ${\rm
    Dec.=-29\degr52\arcmin16\arcsec}$.  The exposure time for the UV
  data is 13.5\,ks, which makes it one of the deepest UV integrations
  on a nearby galaxy to date. The UV maps reveal a rich spiral structure
  within the star forming disk, a diffuse extended envelope of UV
  emission and multiple arms, arcs and filaments that reach out to
  almost 4 optical radii (i.e., roughly the edge of the map). All
  these structures show a stunning degree of correlation with the {\sc
    Hi} map, indicating (qualitatively) a tight spatial correlation of
  atomic hydrogen and UV emission far into the outer disk of M83.}
\label{fig1}
\end{figure*}

\newpage

\section{Data \& Methodology}
\label{data}

We use \hi\ data from the NRAO\footnote{The National Radio Astronomy
  Observatory is a facility of the National Science Foundation
  operated under cooperative agreement by Associated Universities,
  Inc.} VLA obtained as part of THINGS to construct a map of
\hi\ 21\,cm emission across a $\sim50\arcmin$ field-of-view (Figure
\ref{fig1}, lower left panel).  This map has been corrected for
primary beam attenuation (the VLA primary beam FWHM is $\sim30\arcmin$). To
maximize sensitivity for extended emission, we apply the ``natural"
weighting scheme, which yields a sensitivity (1$\sigma$ RMS) of
$\sim0.8$\,mJy/beam (or $\sim0.2\,{\rm M}_{\sun}{\rm pc}^{-2}$) and a
resolution of $\sim13\arcsec$. M83 is known to contain a significant
amount of its total \hi\ content in the form of diffuse, low column
density \hi\ in its outer disk \citep[e.g.,][]{huchtmeier81}, which
our interferometric observations will not pick up by design. Because
we are focussing on \hi\ emission associated with (localized) star
formation, the potentially missing flux on very large ($\gtrsim
15\arcmin$) scales is not a concern in the current study. When quoting
\hi\ surface densities we include a factor of 1.36 to account for
helium and heavier elements.

The GALEX FUV data are part of our GI program GI3-050.  Due to two
bright stars near M83, the GALEX observations were carried out in petal
pattern mode allowing for the field-of-view to be centered on the
galaxy. The GALEX field-of-view is slightly larger than the area
covered by the \hi\ map. We thus restrict our analysis to the
$\sim50\arcmin$ field covered by both maps. M83 was observed for 9
orbits ($\sim13.5$\,ks) yielding a FUV map about three times as
sensitive as the GALEX Nearby Galaxy Survey M83 map
\citep{gildepaz07a}. We use a gaussian kernel to degrade the
resolution of the FUV map to $13\arcsec$ to match that of the
\hi\ map. In this map, the (median based) $1\sigma$ RMS scatter of the
noise is about $2\times10^{-6}$\,mJy arcsec$^{-2}$.

We adopt inclination ($24\arcdeg$) and position angle ($225\arcdeg$)
from \citet{tilanus93}, distance (${\rm D}\approx4.5\,{\rm Mpc}$) from
\citet{karachentsev04} and the optical radius (${\rm
  r_{25}}\approx7.7\arcmin$) from LEDA \citep{paturel03}. For details
of data processing, the conversion of measured intensities to physical
units (e.g., FUV intensities into estimated SFR surface densities),
assessment of uncertainties, etc., we refer the reader to
\citet{bigiel10}.

\section{Location of FUV Relative to {\sc HI}}
\label{sec:location}

\begin{figure*}
\epsscale{0.6}
\plotone{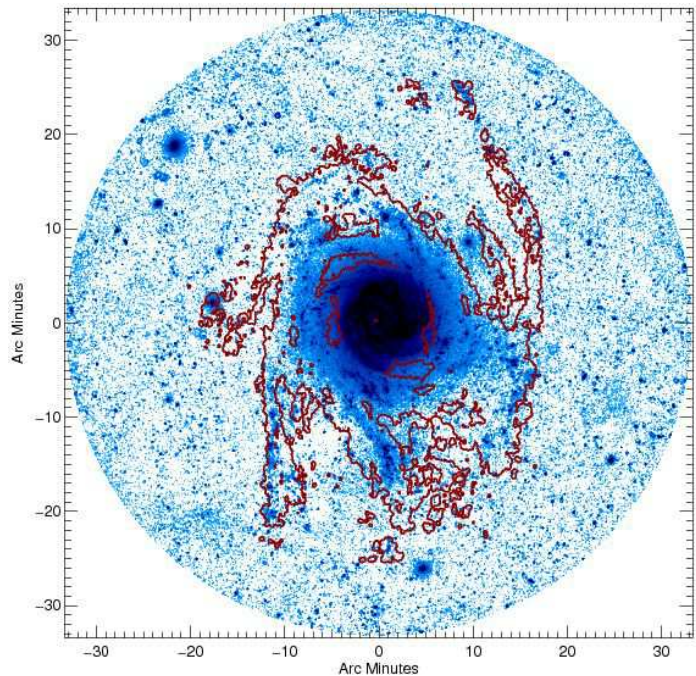}
\caption{Smoothed {\sc Hi} contours of $0.2\,{\rm M}_{\odot}~{\rm
    pc}^{-2}$ overplotted on the GALEX FUV map (compare Figure
  \ref{fig1}, upper left panel).  The figure illustrates the close spatial
   correspondence of {\sc Hi} and FUV emission along the arms and
      filaments in the outer disk of M83.}
\label{fig5}
\end{figure*}

Figure \ref{fig1} shows that \hi\ (lower left panel) and UV emission
(top panels) are detected far into the outer disk of M83 (the edge of
the maps corresponds to $\sim3.5\,{\rm r_{25}}$).  It also shows a
remarkable spatial coincidence of major features found in the
maps. Figure \ref{fig5} shows this even more clearly: Plotted are
(smoothed) \hi\ contours of $0.2\,{\rm M}_{\odot}~{\rm pc}^{-2}$
($\sim2.5\times10^{19}\,{\rm atoms\,cm}^{-2}$) on top of the GALEX FUV
map. Almost every \hi\ feature has a corresponding feature in the FUV
map. Even very remote structures, such as the tip of the extended,
western \hi\ arm show FUV emission and thus signs of recent star
formation activity. This close correspondence implies that the
extended \hi\ disk represents the reservoir from which the young stars
emitting the FUV emission are forming (presumably with an intermediate
stage of bound molecular gas). It also implies, quite surprisingly,
that massive star formation proceeds -- if at a low level -- almost
everywhere \hi\ is observed.

\begin{figure*}
\epsscale{0.5}
\plotone{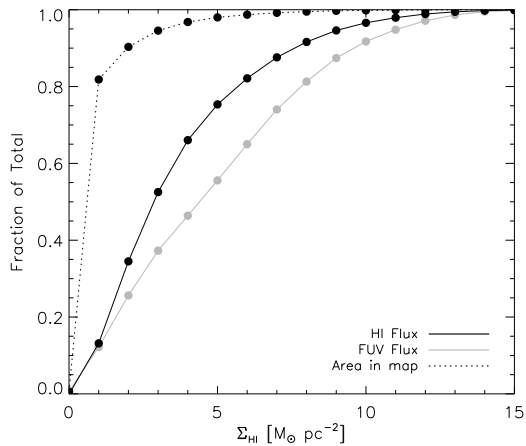}
\caption{The distribution of FUV flux (gray), {\sc Hi} flux (black) and area (dotted line) in our maps as a function of {\sc Hi} surface density. The $y$-axis shows the fraction of the total
  {\sc Hi} or FUV flux or the fraction of area (relative to the entire map) below a given {\sc Hi} surface density ($x$-axis). Even though a large fraction of the M83 map is covered by little
  or no {\sc Hi} (dotted line), most of the FUV light comes from regions with relatively high {\sc Hi} surface densities (gray line).}
\label{fig3}
\end{figure*}

Figure \ref{fig3} quantifies the coincidence of \hi\ and FUV\footnote{For this plot (and the
  following) we exclude regions near foreground stars and background galaxies identified from
  the GALEX pipeline source catalogs and from comparing UV and optical imaging}. The
$y$-axis gives the fraction of the total FUV (gray) or \hi\ (black)
flux found in regions (i.e., along lines-of-sight) with \hi\ surface densities lower than the value on the
$x$-axis. For example, for an $x$-axis \hi\ surface density of
2~M$_\odot$~pc$^{-2}$, we identify all area (i.e., lines-of-sight) in the maps with $\Sigma_{\rm
  HI} < 2$~M$_\odot$~pc$^{-2}$. We then add up the \hi\ (FUV) emission from this area and
divide this \hi\ (FUV) flux by the \hi\ (FUV) flux of the entire map to get the fraction of
flux found below the \hi\ surface density of 2~M$_\odot$~pc$^{-2}$. For comparison, we also plot
(dotted line) the fraction of area as a function of \hi\ surface
density (e.g., in the above example the fraction of the entire M83 \hi\ map with
$\Sigma_{\rm HI}< 2$~M$_\odot$~pc$^{-2}$).

The key point in Figure \ref{fig3} is that the FUV curve (gray) tracks
the \hi\ curve (black) much more closely than it follows the area
curve (dotted). The area curve is the expected distribution
for FUV emission if it were randomly distributed across the map. The
fact that the \hi\ and FUV curves are quite similar thus shows that rather than being
randomly distributed, the distribution of FUV flux seems to follow that of the
\hi . Also, while most of the area in our map is covered by low column (or no)
\hi, the FUV comes mostly from lines-of-sight with
relatively high $\Sigma_{\rm HI}$. For example, $90\%$ of the area has
\hi\ surface densities $<2{\rm M}_{\odot}~{\rm pc}^{-2}$, while only
$\sim25\%$ of the FUV flux comes from this area.  This behavior is
expected if FUV is spatially correlated with \hi\ emission and implies
that a substantial amount of \hi\ (high surface densities) seems to be
a necessary prerequisite to star formation.

\section{Radial Trends}
\label{sec:amount}
\begin{figure*}
\epsscale{1.0}
\plottwo{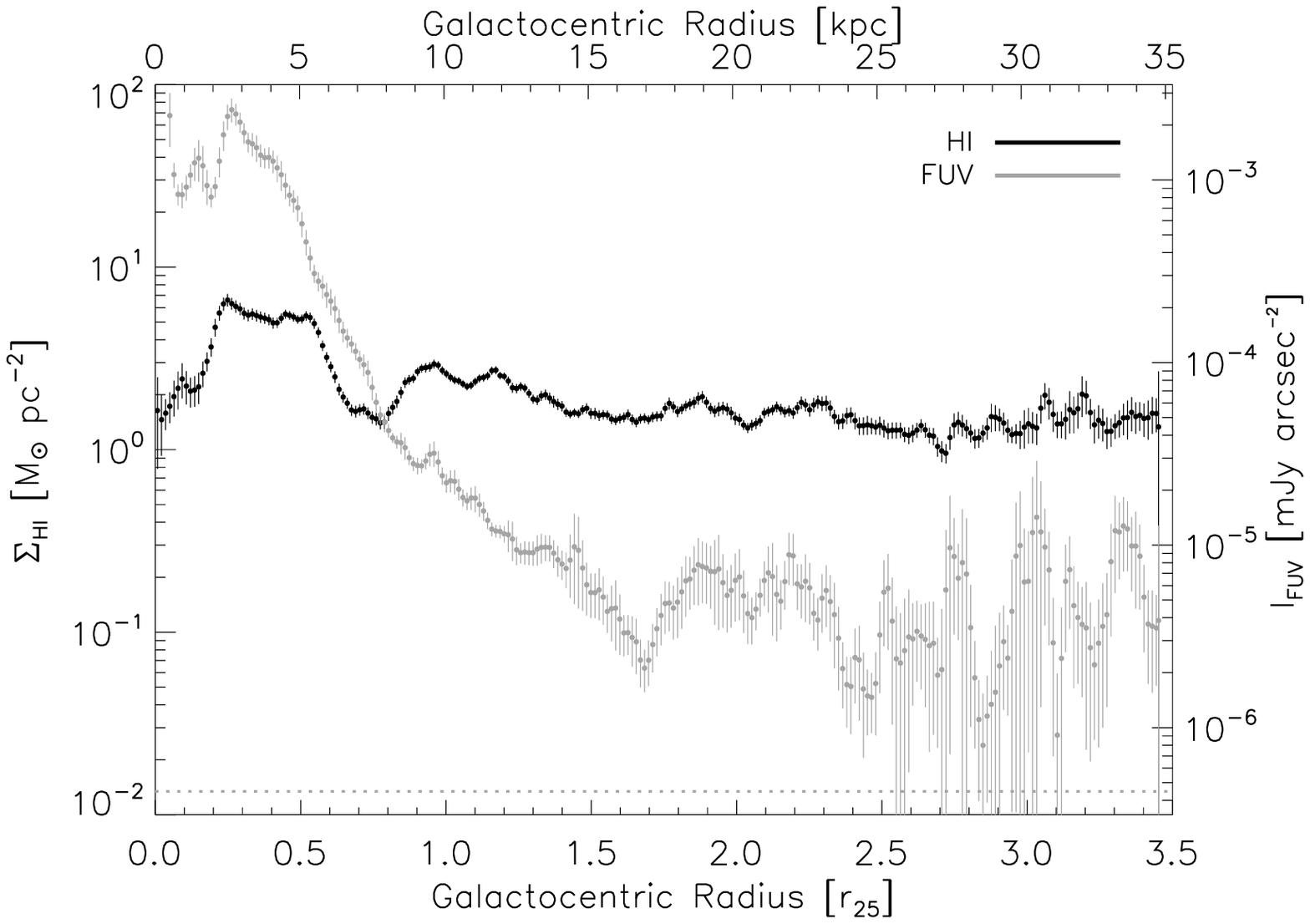}{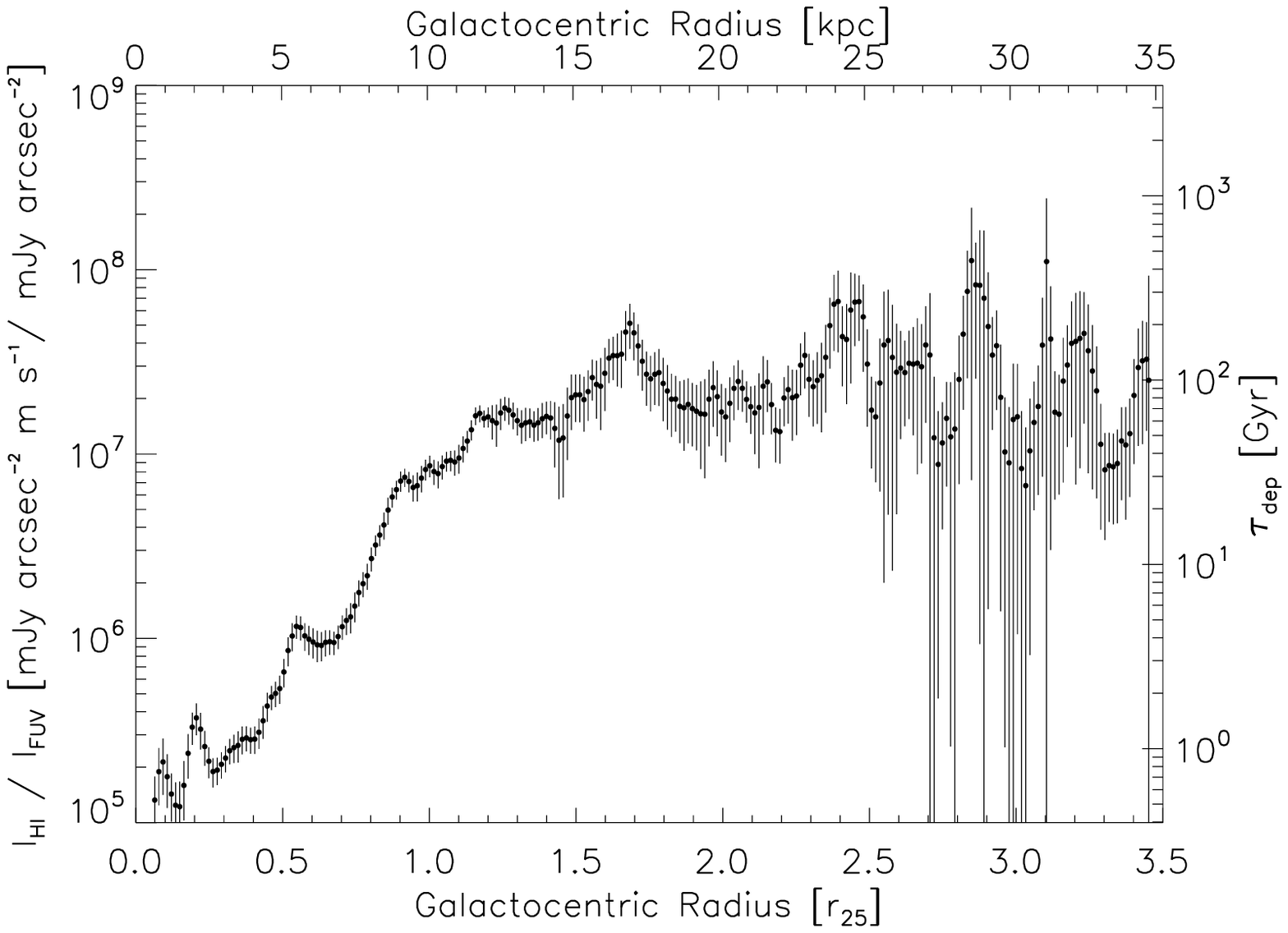}
\caption{{\em Left:} {\sc Hi} (black) and FUV (gray) emission
  averaged in $6.5\arcsec$ wide azimuthal rings
  (deprojected radial profiles) in regions of significant {\sc Hi} emission, i.e., along the
  arms in the extended disk of M83 (see text). The error bars show the ($1\sigma$) uncertainty
  in the mean in each annulus. The left axis provides the mass surface
  density scale for the {\sc Hi} profile, the right axis the intensity
  scale for the FUV profile (we note that in order to convert the FUV intensity scale -- in units of
  ${\rm mJy\,arcsec^{-2}}$ -- into SFR surface densities $\Sigma_{\rm SFR}$ -- in units of
  ${\rm M}_{\sun}{\rm yr^{-1}}{\rm kpc^{-2}}$ -- one needs to multiply
  $I_{\rm FUV}$ by $\sim3.46$; compare \citet{bigiel10}). The gray dotted line indicates the typical
  $5\sigma$ sensitivity for the (averaged) FUV emission in an (outer disk)
  annulus. The respective {\sc Hi} sensitivity is below the lower
  plot limit. {\sc Hi} and FUV emission show quite distinct radial
  trends: whereas average {\sc Hi} surface densities remain relatively
  constant along the filaments in the outer disk, the mean FUV
  intensity continues to drop before leveling off at $1.7{\rm
    r_{25}}$. {\em Right:} Intensity ratio of {\sc Hi} and FUV radial
  profiles. Converting FUV intensity into $\Sigma_{\rm SFR}$, this ratio yields the
  {\sc Hi} depletion time (right axis). At large radii, this depletion time
  remains relatively constant at about 100\,Gyr, i.e., many Hubble
  times.}
\label{fig4}
\end{figure*}

Having seen that FUV and \hi\ emission are remarkably coincident, we
now ask how the {\em amount} of FUV emission compares to the mass of
\hi\ along a line-of-sight. The \hi\ is concentrated into features
(the arms and filaments in Figures \ref{fig1} and \ref{fig5}) with a
relatively low filling fraction, and we have just seen that FUV
emission is largely coincident with these features. Our approach here
is to look at the mean ratio of \hi-to-FUV {\em along these features}
as a function of radius, asking ``Where there is \hi, what is its
average surface density and what is the associated average FUV
intensity?".  For this, we consider only emission from areas that have
\hi\ surface densities (in a smoothed 30$\arcsec$ resolution version
of the \hi\ map) $\gtrsim0.5\,{\rm M}_{\sun}{\rm pc}^{-2}$.

Figure \ref{fig4} shows the average \hi\ and FUV intensities along
these features. Differences are immediately apparent: the FUV profile
(gray) decreases over many orders-of-magnitude from the inner parts
into the outer disk before it levels off at about $1.7{\rm r_{25}}$
(note that the FUV emission is not corrected for internal extinction,
which is significant at small radii). The FUV emission at large radii
is still detected with good sensitivity; the profile is always well
above the gray dotted line, which shows the typical 5$\sigma$
sensitivity. This supports the qualitative conclusion from Figure
\ref{fig5} that the extended \hi\ structures in M83 seem to coincide
with FUV emission, indicating recent SF activity
out to largest radii. In contrast to the FUV profile, the \hi\ surface
density (black) shows only a moderate change with radius. The typical
(azimuthally averaged) \hi\ surface density is $\sim2-4\,{\rm M}_{\sun}{\rm
  pc}^{-2}$.

Assuming FUV emission to arise mostly from young massive stars forming
at a steady rate, we can convert the measured FUV intensities into star formation
rate surface densities, $\Sigma_{\rm SFR}$ \citep[using the same
  procedure as][]{bigiel10}. Then the ratio of the two profiles
$\Sigma_{\rm HI}/\Sigma_{\rm SFR}$ (which is proportional to $\Sigma_{\rm
  HI}/I_{\rm FUV}$) is the \hi\ depletion time, the time it takes
(present day) SF to deplete the current supply of \hi. This quantity
is shown as a function of radius in the right panel of Figure
\ref{fig4} (note that inside $\sim 0.5{\rm r_{25}}$ we do not expect
$\Sigma_{\rm HI}/I_{\rm FUV}$ to trace the gas depletion time, as both
\htwo\ and extinction become important).

The \hi\ depletion time rises from the optical disk ${\rm r_{25}}$ out to
$\sim1.7{\rm r_{25}}$ and then remains relatively constant across the
far outer disk of M83 at about 100\,Gyr, i.e., many Hubble times. This
very long timescale has several implications. First, it is much longer
than the $\sim 2$\,Gyr depletion time of the {\em molecular} gas observed
in the inner disks of galaxies \citep{bigiel08,leroy08}. Assuming bound, molecular clouds
form stars with roughly the same efficiency in the inner and outer
disk, this extremely long \hi\ depletion time suggests that the process
of building these molecular clouds is the bottleneck for
forming stars at large radii. Combined with the rather low
\hi\ columns that dominate at large radii, the regulating and / or
limiting factor for star formation in these environments may well be
that of driving the \hi\ to high enough densities to trigger
the atomic-to-molecular phase transition and subsequent star formation.

The long \hi\ depletion time also implies that the extended
\hi\ reservoir in the outer disk is long-lived, or at least that {\em
  in situ} star formation is unlikely to consume the gas. This means
that this gas is in principle available as fuel for star formation in
the inner disk, if it can be efficiently transported into the inner part
of the galaxy \citep[e.g.,][]{wong04,vollmer10}. Single dish observations sensitive
to the entire \hi\ distribution (compare \S\,\ref{data}) suggest that M83 hosts a very large reservoir of
\hi\ beyond the optical disk -- as much as 80\% of the total \hi\ mass
\citep[][and references therein]{huchtmeier81}. The total mass of this
\hi\ reservoir may well exceed $\sim10^{10}{\rm M}_{\sun}$, which
would in principle be enough to sustain present-day SF in the inner
disk of M83 for 2--3 times longer than is possible from the
\htwo\ alone \citep{lundgren04}.

The moderate SF activity at large radii should also chemically enrich
the \hi\ envelope. If the disk remained in its present configuration
for $\sim 5$~Gyr, then the integrated outer-disk SFR ($\sim0.05\,{\rm
  M}_{\sun}{\rm yr^{-1}}$ beyond ${\rm r_{25}}$) and a stellar yield
of $\sim0.06$ \citep{krumholz10} imply a mass of heavy elements equal
to about $1\%$ of the total outer disk \hi\ mass (as determined from our map). This is only a
coarse estimate, but it shows that even the low observed outer disk SFR may
enrich the outer disk gas to a moderate metallicity. This fits to
observations of emission from hot dust \citep{dong08} and
fairly decent metallicities \citep{gildepaz07b,bresolin09} in the XUV disk of
M83.

\section{Distribution of {\sc HI} Surface Densities}
\label{dist-hi}
\begin{figure*}
\plottwo{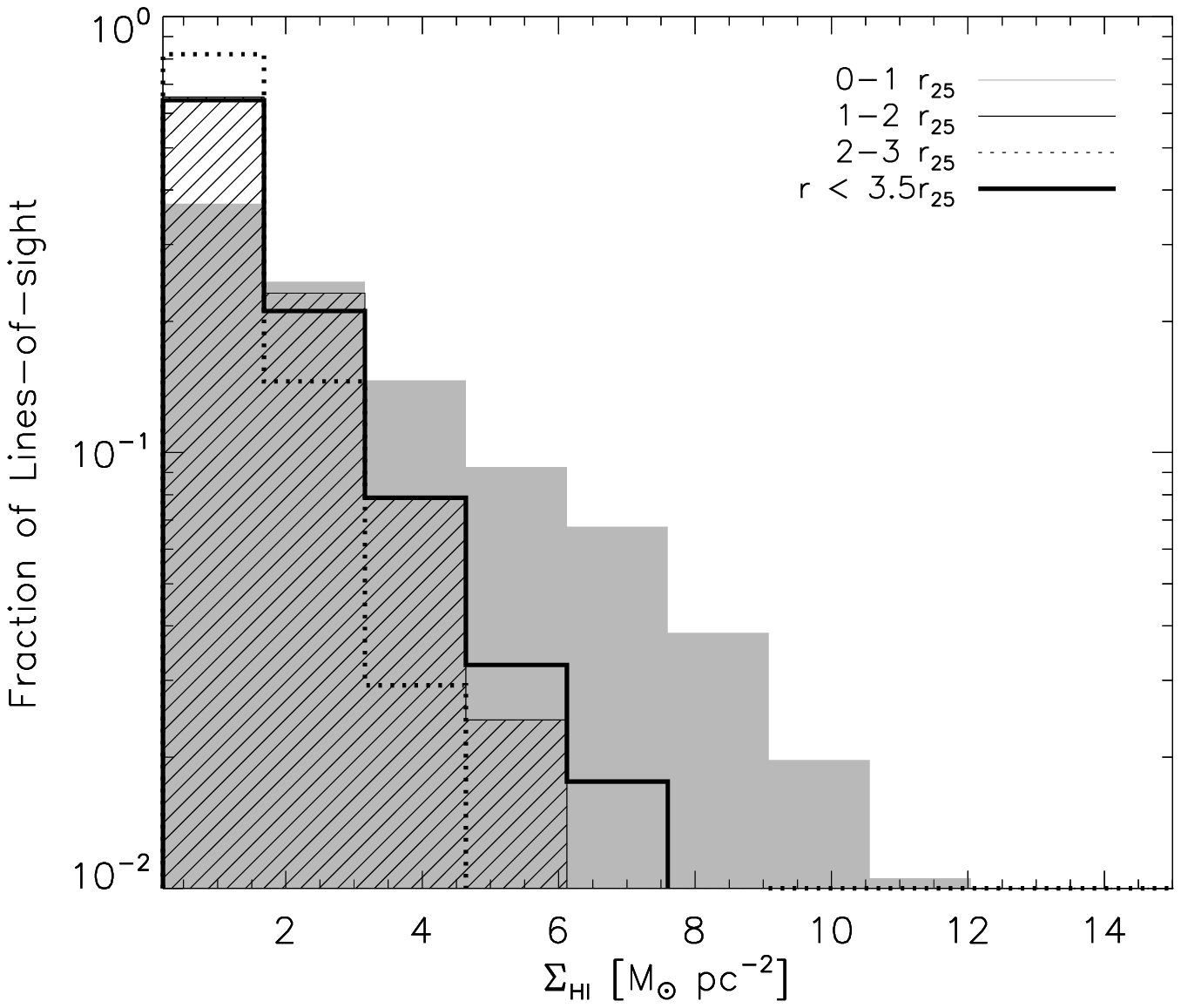}{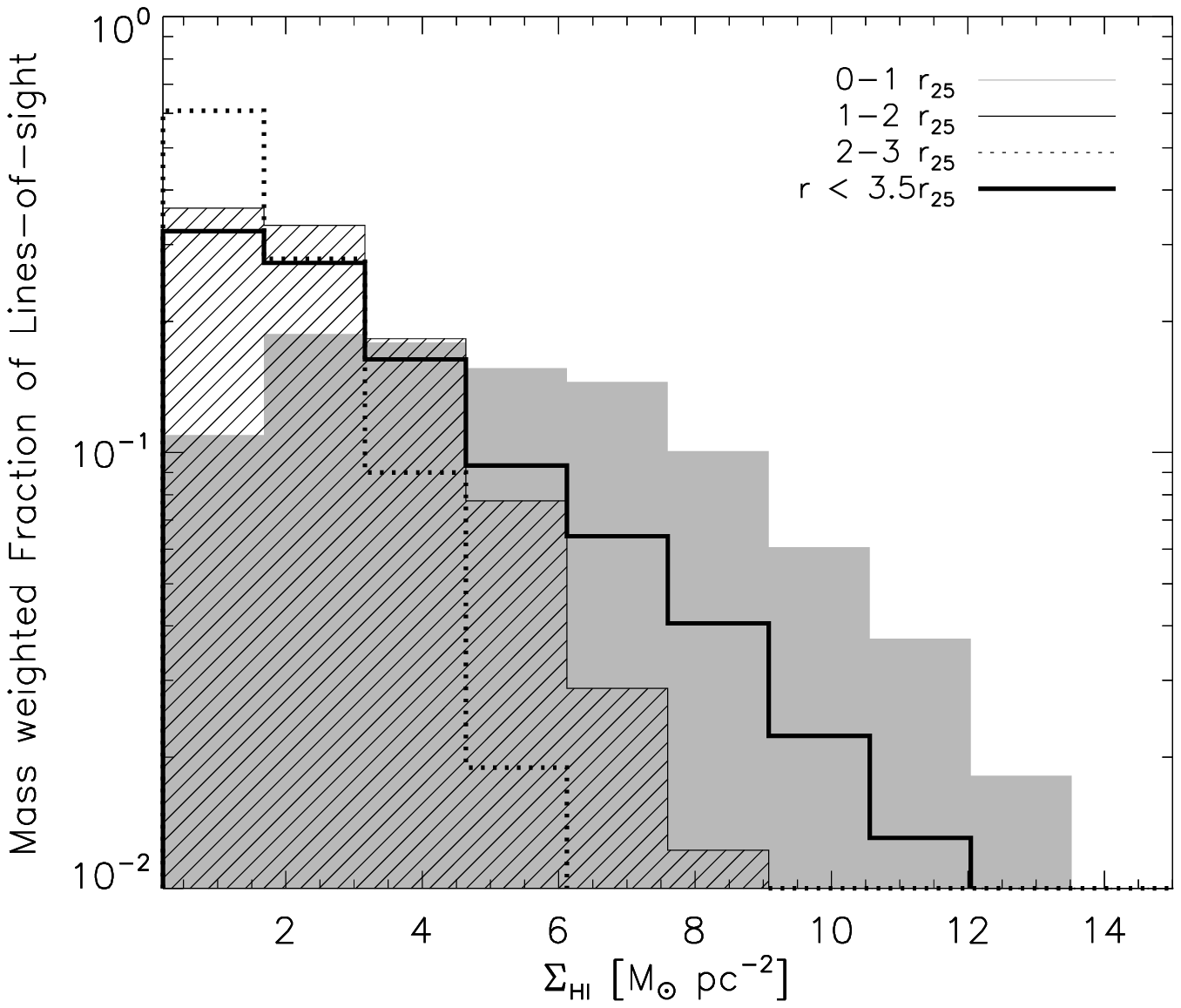}
\caption{Normalized histograms of {\sc Hi} surface density (left
  panel) -- the right panel shows the same distribution but mass-weighted -- for each line-of-sight in 3
  radial regimes: $0-1\,{\rm r_{25}}$ (gray), $1-2\,{\rm r_{25}}$
  (black hashed), $2-3\,{\rm r_{25}}$ (black dotted) and for the
  entire {\sc Hi} disk of M83 (thick black line). We bin emission
  between 0.2 (the sensitivity of the {\sc Hi} map) and $15\,{\rm
    M}_{\odot}~{\rm pc}^{-2}$ in equally sized bins. The plots show
  that low {\sc Hi} columns dominate most of the area at all radii in
  M83 and that while higher columns contribute most to the mass within
  the optical disk, low column {\sc Hi} dominates the mass budget in
  the outer disk.}
\label{fig2}
\end{figure*}

We have seen that low level star formation is pervasive in the
\hi\ arms and filaments in the outer disk of M83. We have also seen that
the absolute level of this star formation is very low compared to that
found in galaxy disks. What drives this drop in the fraction of
\hi\ that eventually ends up in stars? A number of authors have suggested that
a drop in \hi\ column or volume density leads to some critical
prerequisite for cloud formation not being met
\citep[e.g.,][]{elmegreen94,schaye04,bush10}. We conclude by briefly
examining how \hi\ surface densities actually change between inner and outer
disk.

The left panel in Figure \ref{fig2} shows the (normalized)
distribution of $\Sigma_{\rm HI}$ for lines-of-sight in
different radial regimes (see caption for details). The right panel
shows the same data, but now each line-of-sight is weighted by mass, so that the left panel
gives the distribution of surface densities and the right panel shows
at what column most of the mass is found.

In the left panel, the histograms from different regimes are not
particulary distinct though one can see some enhancement in the
\hi\ columns found inside $r_{25}$ (visible also as the ``bump" in the
radial profile in the left panel of Figure \ref{fig4} between $\sim0.2$ and
$\sim0.7\,{\rm r_{25}}$).  These differences are much easier to see in
the mass-weighted histograms on the right.  While in the optical disk
\hi\ surface densities between $\sim2-8\,{\rm M}_{\odot}~{\rm
  pc}^{-2}$ contribute roughly equally to the integrated \hi\ mass in
this regime, it is clearly the lowest \hi\ columns that contribute
most to the mass in the outer disk (i.e., beyond ${\rm r_{25}}$). In
fact, there appears to be a radial trend in the sense that as one
moves inwards from the far outer disk, an increasing fraction of the
\hi\ mass comes from lines-of-sight with relatively high \hi\ surface density.

We saw that in the outer disk of M83, high \hi\ surface densities, likely a prerequisite for
molecular gas and subsequent star formation, contribute a decreasing
fraction to the overall \hi\ mass as one moves to large radii. The
extended arms seen in Figure \ref{fig1} thus seem to represent merely the ``tip of
the iceberg" of the extended \hi\ distribution in M83. These arms, however, appear to be
the regions where pervasive star formation can proceed even at extreme galactocentric radii.

\acknowledgments
F.B. acknowledges support from NSF grant AST-0838258 and earlier support from the Deutsche
Forschungsgemeinschaft (DFG) Priority Program 1177. Support for A.L. was provided by NASA through Hubble Fellowship grant
HST-HF-51258.01-A awarded by the Space Telescope Science Institute, which
is operated by the Association of Universities for Research in Astronomy,
Inc., for NASA, under contract NAS 5-26555. We have made use
of the NASA/IPAC Extragalactic Database (NED), which is operated by the Jet
Propulsion Laboratory, California Institute of Technology, under
contract with the National Aeronautics and Space Administration. This
research has made use of NASA's Astrophysics Data System (ADS).
We acknowledge the usage of the HyperLeda database (http://leda.univ-lyon1.fr).

\end{document}